\documentclass[journal,table]{IEEEtran}
\usepackage{scrtime}
\usepackage[dont-mess-around]{fnpct}

\usepackage{etoolbox}
\makeatletter
\patchcmd\maketitle{\setcounter{footnote}{0}}{}{}{}
\patchcmd\maketitle{%
  \renewcommand\thefootnote{\@fnsymbol\c@footnote}}{\AdaptNote\thanks\multthanks}{}{}
\patchcmd\maketitle{%
  \def\@makefnmark{\rlap{\@textsuperscript{\normalfont\@thefnmark}}}}{}{}{}
\makeatother
\usepackage{blindtext}
\usepackage{graphicx}
\usepackage{ulem}
\bstctlcite{IEEEexample:BSTcontrol}
\usepackage[T1]{fontenc}
\usepackage{relsize}
\usepackage{amsthm}
\usepackage{mathtools}
\usepackage{cuted}
\usepackage{multicol}
\usepackage{amsmath,nccmath}

\usepackage[cmintegrals]{newtxmath}
\usepackage{xcolor}
\usepackage[utf8]{inputenc}
\usepackage[T1]{fontenc}
\usepackage{placeins}
\usepackage{tabularx}
\usepackage[T1]{fontenc}
\usepackage{float}
\usepackage{algpseudocode}
\usepackage[vlined,ruled]{algorithm2e}
\include{pythonlisting}
\definecolor{algoColorKeyword}{named}{blue}
\definecolor{algoColorComment}{named}{green}

\makeatletter

\makeatletter
\def\blfootnote{\xdef\@thefnmark{}\@footnotetext}
\makeatother

\let\labelindent\relax
\usepackage{enumitem}
\usepackage{amsfonts}
\usepackage{alltt}
\usepackage{graphicx}
\usepackage{url}
\usepackage{tabularx}
\usepackage{tabulary}
\usepackage{listings}
\usepackage{epstopdf}
\usepackage{makecell}
\usepackage{graphicx}
\usepackage{caption}
\usepackage[caption=false]{subfig}

\SetCommentSty{mycommfont}
\begin{document}

\setlength{\abovedisplayskip}{0pt}
\setlength{\belowdisplayskip}{0pt}
\title{CPVNF:Cost-efficient Proactive VNF Placement and Chaining for Value-Added Services in Content Delivery Networks}

\author{Mouhamad Dieye,~\IEEEmembership{Member,~IEEE,}
Shohreh Ahvar,~\IEEEmembership{Member,~IEEE,}
Jagruti Sahoo,~\IEEEmembership{Member,~IEEE,}
Ehsan Ahvar,~\IEEEmembership{Member,~IEEE,}
Roch Glitho,~\IEEEmembership{Senior Member,~IEEE,}
       Halima Elbiaze,~\IEEEmembership{Member,~IEEE,}
       Noel Crespi,~\IEEEmembership{Senior Member,~IEEE,}
}

\markboth{}%
{Shell \MakeLowercase{\textit{et al.}}: Bare Demo of IEEEtran.cls for Journals}

\maketitle

\begin{abstract}
Value-added services (e.g., overlaid video advertisements) have become an integral part of today's Content Delivery Networks (CDNs). To offer cost-efficient, scalable and more agile provisioning of new value-added services in CDNs, Network Functions Virtualization (NFV) paradigm may be leveraged to allow implementation of fine-grained services as a chain of Virtual Network Functions (VNFs) to be placed in CDN. The manner in which these chains are placed is critical as it both affects the quality of service (QoS) and provider cost. The problem is however, very challenging due to the specifics of the chains (e.g., one of their end-points is not known prior to the placement). We formulate it as an Integer Linear Program (ILP) and propose a cost efficient Proactive VNF placement and chaining (CPVNF) algorithm. The objective is to find the optimal number of VNFs along with their locations in such a manner that the cost is minimized while QoS is met. Apart from cost minimization, the support for large-scale CDNs with a large number of servers and end-users is an important feature of the proposed algorithm. Through simulations, the algorithm's behavior for small-scale to large-scale CDN networks is analyzed.
\blfootnote{
\noindent Mouhamad Dieye and Halima Elbiaze are with Université du Québec À Montréal.\\
Roch Glitho is with Concordia University, Montreal, QC H3G 1M8, Canada,
and also with the University of Western Cape, Cape Town 7535, South Africa.\\
Shohreh Ahvar and Noel Crespi are with Institut Mines-Télécom, France\\
Jagruti Sahoo is with South Carolina State University, USA\\
Ehsan Ahvar is with Université Rennes, INRIA, CNRS, IRISA, France\\
}
\end{abstract}


\IEEEpeerreviewmaketitle

\section{Introduction}
Content Delivery Networks (CDNs) are largely distributed infrastructures of surrogate servers placed in strategic locations \cite{Pallis:2006} \cite{Pathan2008}. Content is replicated on these servers in order to serve end-users with reduced latency. Beyond hosting content, the popularity of CDNs as a platform for delivering value-added services has increased over the years. Many CDN providers such as Akamai, Limelight, etc. offer value-added services. Currently around 47\% of Akamai’s revenues come from its value-added service offerings, which also provide higher margins compared to its basic CDN services \cite{Akamai}.
Typical value-added services provided by CDNs include website/application acceleration (e.g., route optimization, TCP optimization, stream splitting) \cite{Pathan2014}, analytics, content protection, advertisement overlays/tickers and content adaptation (e.g., transcoding, compression). Website/application acceleration relies on a combination of optimization techniques and allows end-users (mostly enterprises) to access a website/application with improved response time. CDN analytics services provide awareness at various levels including network, device and content, thereby enabling content providers to make more informed and business-friendly decisions. 
Increasingly, multimedia content providers are finding it useful to outsource their media-related services such as digital rights management and content adaptation to the CDN provider domain. This is because providing such services requires the content provider 
to store customized content for every end-user. The overlay/ticker service is perceived as a fast monetization strategy by content providers with, for instance 
local news/weather updates
inserted on the top of the video as static or scrolling tickers. Similarly, short video advertisements are linearly overlaid on the video delivered to end-users.
Traditional CDNs face numerous obstacles towards efficiently provisioning value-added services. First, the value-added services are deployed on a dedicated hardware in the CDN infrastructure. As a result, it is both time-consuming and expensive to deploy and manage the service, resulting in more time to market and cost-inefficiency. Second, the explosive growth in end-users and amount of content \cite{Index} delivered to them raise the need to scale the deployed services as needed. Third, end-users' growing interest in customized content fueled by a continual innovation in video formats over the years requires rapid provisioning of novel video-based value-added services. 
Network Functions Virtualization (NFV) \cite{Han} \cite{Mijumbi} is an emerging paradigm that can be used to make CDNs meet the above-mentioned requirements. NFV is a novel telecommunication service provisioning approach in which the network function is decoupled from the physical devices on which they run and are implemented as virtualized software, termed as virtual network functions (VNFs) which run on top of a virtualized infrastructure and chained together to provide a required service. They can be implemented on any computational node (e.g., CDN surrogate server, switches, data center) that meets their resource demand. The computational nodes must provide Network Functions Virtualization Infrastructure (NFVI) functions to support the execution environment of VNFs. These nodes are referred to as NFVI nodes. It should be noted that NFV has been traditionally used to virtualize middle-boxes (e.g., firewall, Network Address Translator (NAT) and Deep Packet Inspection (DPI)) \cite{Martins:2014}. Recently, NFV has been investigated in other domains such as virtualized Wireless Sensor Networks (vWSNs) \cite{Mouradian}, optical networks \cite{Munoz} and IP Mutimedia Subsystems (IMSs) \cite{Carella}.
NFV could enable CDNs to provision value-added services with significantly lower deployment and maintenance costs. Thanks to virtualization techniques, NFV allows the value-added services to scale in an elastic manner. Moreover, due to the dynamic service chaining feature of NFV, the update of an existing value-added service or the introduction of new value-added services could be achieved with increased agility by inserting and/or removing VNFs to and from an existing service chain on-the-fly. We term the CDN architecture that relies on NFV as NFV-based CDN.

The placement and chaining of VNFs
affects the desired QoS (e.g., delay) of a value-added service and the cost of the CDN provider. It is modeled as an optimization problem where the objective is to find the optimal number and location as well as efficient chaining of VNFs instances such that the CDN provider cost is minimized and QoS is satisfied. In this paper, the cost includes the license, computing and communication cost. Where as the license cost includes instances and sites license and is computed based on the number of utilized VNF instances and sites (i.e., the servers). The computing cost includes the cost of running VNFs on servers and the communication cost is defined as the sum of the bandwidth used by the chains in the network. Although, some studies (e.g., \cite{Bari}) consider the cost of transferring, booting and attaching a VM image to devices before deploying a VNF, this research does not consider this cost in the CDN environment as it is assumed that the CDN provider is the NFVI owner.
In this paper, we focus on a proactive placement of VNFs where VNFs are deployed in an optimal manner before any request is received from the end-users to access the service. This type of deployment is triggered when a content provider requests the CDN provider to deploy a set of value-added services. The request specifies the service-related parameters (e.g., the description of functionalities that constitute the services) including the QoS threshold to be satisfied. 

The rest of the paper is organized as follows: The next section illustrates the problem via a use case, introduces the key requirements and reviews related works. Section 3 presents the VNF placement problem in CDNs and ILP formulation. Section 4 describes the proposed algorithm. Section 5 portrays the simulation results. Section 6 concludes the paper and outlines some future works.

\section{ILLUSTRATIVE USE CASE, REQUIREMENTS AND RELATED WORK}
To the best of our knowledge, the only paper to discuss a use case for VNF–based value-added service provisioning in CDN is reference \cite{Jahromi} which focuses on the architectural aspects while we focus on the algorithmic aspects. The European Telecommunications Standards Institute (ETSI) has also proposed a use case on the virtualization of CDNs entities (e.g., surrogate server, CDN controller, etc.) \cite{ETSI}, but with no bearing on value-added service provisioning. Besides, few NFV architectures for CDN have been proposed in the literature \cite{Giotis} \cite{Herbaut}, yet with no bearing on value-added services provisioning. 
\subsection{Illustrative Use Case}\label{illustrationCase}
We assume a business model with the following entities: Content provider (e.g., YouTube), CDN provider (e.g., Limelight), VNF provider and end-users. Content provider provides the value-added services to its end-users. The CDN provider owns surrogate servers and operates NFVI on the surrogate servers. The VNF provider provides VNFs. End-user consumes the value-added service. The reader should note that like in any business model, the same actor may play several roles at the same time.

\begin{figure}[h]
    \centering
    \includegraphics[width=\columnwidth]{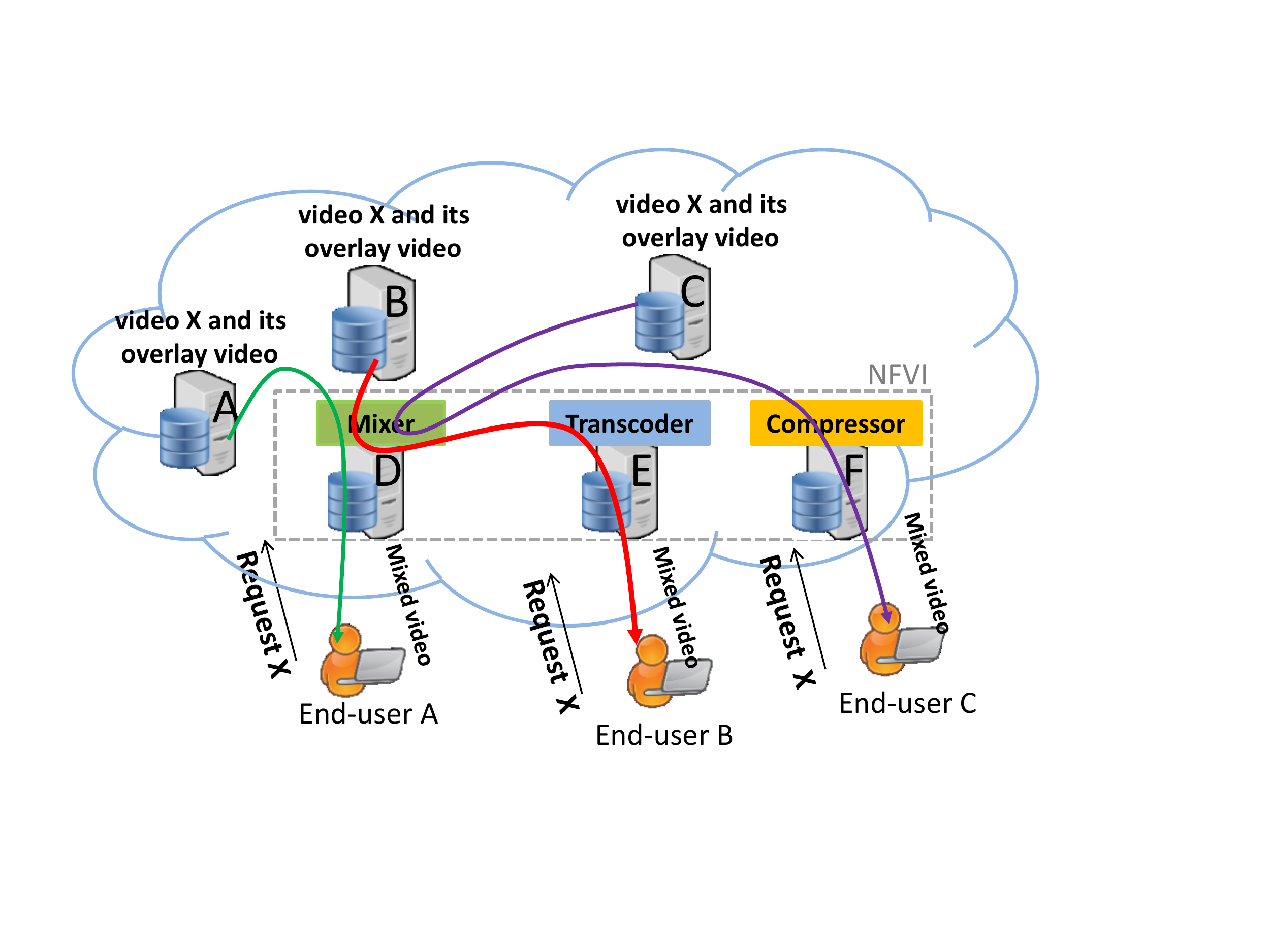}
    \caption{Video Advertisement Overlay Use case}
    \label{fig:my_label}
\end{figure}
Let us consider a simplified scenario as shown in Figure 1. It depicts three end-users A, B and C, with mobile devices of different capabilities (e.g. supported codecs, resolutions, bandwidth and processing power) who wish to access a video X. As per CDN principles, the video is duplicated over several servers (i.e. A, B, and C in this case). When the end-users request the video, the CDN provider adds an overlay video to it as a value-added service. Such an overlaid video enriches the viewing experience of end-users by providing a clip of advertisement based on the end-user’s interest. In this use case, it is assumed that video X and the video which is overlaid on it reside on the same servers. It should be noted that the video overlay service requires a video mixer functionality to embed the advertisement in the  main video.

The end-users may however not be able to play the mixed video on their devices depending on their capabilities. In this scenario, it is only end-user A who is able to play the mixed video. End-user B does not have the required codec, therefore the video must goes through a transcoder before being delivered to her/him. As for end-user C she/he has a limited bandwidth capacity, and therefore, its device cannot play the video in its original resolution. Consequently, the mixed video needs to be compressed before delivery. Three fine-grained functionalities are therefore required this value added service:  mixer, transcoder and compressor. 
Their deployment starts when the content provider sends a request to the CDN provider to deploy the video value-added services. The CDN provider then gets them from the VNF provider and uses a VNF placement algorithm to deploy them on a selected subset of surrogate servers to optimize given criteria (e.g. QoS) delay when the video is viewed. The reader should note that surrogate A, B, and C which will serve end-user A, B and C, respectively are not known when the deployment is made.  They will be selected dynamically by the CDN controller server when the end-users will request the services. 
\subsection{Requirements}
Achieving optimal placement of VNFs is an issue in provisioning value-added services in CDNs because of the inherent goals associated. First, the services must be delivered to end-users with high QoS (e.g., reduced service delay). It is in fact desirable to provide the service with a strict QoS guarantee. Second, the service placement should be operational in small and large scale CDNs. Third, the cost incurred in deploying services and delivering them to the end-users must be minimized for cost-efficiency. 
Thus, the optimal placement of VNFs in CDN consists of determining how many VNF instances are required to meet the end-user workload and on which surrogate servers they should be deployed so that the above-mentioned goals are reached. The placement of VNFs in CDN can be achieved in two ways: 1) Proactive, and 2) Reactive. This paper focuses on the proactive placement.
The general requirements for the proactive placement are summarized as follows:
\begin{itemize}
    \item To ensure QoS, especially in terms of service delay
    \item To ensure low cost
    \item To be operational in small and large scale CDNs 
\end{itemize}

\subsection{Related Work}
Although VNF placement algorithms constitute a large area of interest for many researchers, very few of them have tackled the problem of placing VNFs in CDNs. Here, we first summarize the main research results on VNF placement in general and also on VNF placement in CDNs. We then show why these results are not applicable to our specific problem of placing VNFs for VAS in CDNs.

\subsubsection{VNF placement algorithms in general}

VNF placement algorithms constitute a large area of interest for many researchers. Here, we first summarize the main results in the area, and then, show why they are not applicable to the CDN context. 

Most VNF placement algorithms deal with cost as an optimization objective. A VNF cost is generally made up of a set of individual costs (e.g., instance license, site license, deployment and communication cost). Some of the existing work focus on specific individual costs while others focus on a set of individual costs. They are discussed below. 

\textbf{Algorithms with single costs as objective\\}
Luizelli et al. \cite{Luizelli} propose an ILP model to minimize the number of instances in order to reduce license costs. They further propose a binary search-based algorithm to improve the ILP run-time. Fang et al. \cite{Fang} also attempt to minimize the number of deployed instances by proposing an ILP and the longest common sub-sequence (LCS)-based heuristic in inter-datacenter elastic optical networks (inter-DC EONs). They also consider the spectrum utilization cost for fiber links, which is the specialized cost for optical network.
Moens et al. \cite{Moens} present an ILP to minimize the number of used servers (or compute resources) for the resource allocation of VNFs in NFVI and also for hybrid infrastructures where some NFs are virtualized and others use specific hardware appliances.

Some other studies have mainly focused on the communication cost. Qu et al. \cite{Qu} propose an ILP and a greedy shortest-path-based heuristic to construct chains through highly reliable VNFs in the NFV-enabled enterprise datacenter networks with the goal of minimizing the communication bandwidth usage across the network. Xia et al. \cite{Xia} formulate the problem in binary integer programming (BIP) and propose a heuristic with the goal of minimizing the overall optical-electrical-optical (O/E/O) conversions (inter-DC traffic) in packet/optical DCs.

\textbf{Algorithms for multiple costs\\}
Unlike the above-mentioned works with the simple objective, the following studies have considered more complex cost models. Ghaznavi et al. \cite{Ghaznavi} present a solution called Simple Lazy Facility Location (SLFL) to optimize the placement of the same-type VNF instances in response to the on-demand workload. In this study, the elasticity overhead and the trade-off between bandwidth and host resource consumption are considered together. In another study, Ghaznavi et al. \cite{Ghaznavi2} propose a Mixed Integer Programming (MIP) model and a heuristic called Kariz for multiple VNF instances placement to provide the functionality of a middlebox. Mechtri et al. \cite{Mechtri} provide decomposition-based approach for the placement of virtual and physical network functions chains to maximize the provider's revenue based on the number of accepted CPU and bandwidth resources.
Riggio et al. \cite{Riggio} propose a VNF placement scheme to minimize the links and nodes utilization to increase the accepted service chain requests in enterprise WLANs. The authors then have extended their work in \cite{Riggio2} where a VNF placement heuristic called WiNE (Wireless Network Embedding) is proposed. 
Sun et al. \cite{Sun} consider cost as the IT resources used for deploying the VNFs and bandwidths cost. They propose an ILP as well as two versions of a heuristic to solve the VNF placement in online and offline manners. The goal in the online heuristic is to maximize the revenue and the goal in the offline version is to minimize the cost. 
Finally, a few studies have attempted to consider more comprehensive cost models. Lin et al. \cite{Lin} present a MILP and Game Theory based VNF placement with the goal of minimizing the cost to deploy NF instances as well as the computing and network cost in optical networks.
Zeng et al. \cite{Zeng} consider the cost of IT resource and spectrum utilization of fiber links as their objective in addition to the cost of VNF deployment (instantiating) in the VNF placement in optical datacenters. They propose a MILP and some heuristics to solve the problem.
Bouet et al. \cite{Bouet} propose an ILP and a centrality-based greedy algorithm to minimize the cost in virtual DPI (vDPI) placement where the cost includes the network cost, the license cost per site and the one per vCPU for VNF instances.
Bari et al. \cite{Bari} propose an ILP and a heuristic to solve the optimal VNF placement by running the Viterbi algorithm. The authors have also considered a penalty cost to be paid to the customer for the service level objective (SLO) violations.

\subsubsection{VNF placement algorithms in CDNs}
Herbaut et al. \cite{Herbaut2017} is an example of works that deal with the algorithmic solutions to the VNF placement in CDNs. It focuses on CDN operators who aim to deploy surrogate servers as VNFs in Internet Service Providers’ (ISP) network. The ISP provides the NFVI Point of Presence (PoP) for the deployment. However it maintains the confidentiality of its network and resources by offering an abstract view to the CDN operator.  The CDN operator just has access to an overlay that connects its end-users to the surrogate servers which run as VNFs in the ISP network. It does not know where exactly these surrogate servers are deployed. A high level Service Level Agreement (SLA) is used for the negotiation of computing and connectivity resources between the CDN operators and the ISP. This SLA is expressed as a service function chain. The paper discusses the collaboration model between the ISP and CDN operator, proposes an ILP formulation for the service chain embedding problem, and also a heuristic to increase problem tractability. 
Ibn-Khedher et al. \cite{Ibn-Kheder2017} is another example. It proposes OPAC, an optimal placement algorithm for virtual CDNs. An exact algorithm is proposed and evaluated. The algorithm takes as an input the topology of the underlying network, and optimally places, and migrates the virtual surrogate servers in order to increase user satisfaction and decrease server and network loads. A few works have addressed 
the architectural solutions to the VNF placement problems in CDNs. Frangoudis et al. \cite{Frangoudis2017} details the design and the implementation of an architecture that allows a telecommunication operator to lease its infrastructure to content providers for the deployment of surrogate servers as VNFs. The focus is on the northbound REST APIs used for the interactions between the content providers and the telecommunication operator. The functional entities of the architecture are also specified. Some examples are the Customer Interface Manager which exposes the actual REST APIs, the Service Orchestrator which coordinates the actual deployment. Algorithmic issues are not explicitly excluded from the scope of the work.

 \subsubsection{Why are previous work not adequate for the problem at hand?}
The previous work is not adequate for the problem at hand for two main reasons. The first reason is that VNF placement in CDN for value-added service provisioning is fundamentally different from the VNF placement as considered so far by researchers, be it generally or in the specific case of CDNs. This due to the fact that  in previous works, each chain  has a distinct pair of end-points (i.e., source and destination) which both are known prior to the VNF placement. In references \cite{Pallis:2006} and \cite{Pathan2008} for instance which deal with the specific case of CDNs, end-users are assigned to specific surrogate servers prior to the placement. In other words the specific surrogate server which will serve any given end-user is known prior to the placement.
However, in this work while the destination (i.e. the end-user) is known, the source (i.e. the surrogate server which will serve the end-user) is not known prior to the VNF placement because it is dynamically selected by the CDN controller after the VNF placement, Going back to the illustrative use case, while the locations of end-users A, B, C are known as the destinations , there is no way to know prior to the placement that surrogate server A will be ultimately selected by the CDN controller to serve end-user A. The same applies to the choice of surrogate server B for end-user B, and surrogate server C for end-user C. This brings unique challenges to the service
chain placement in CDNs. To the best of our knowledge, our work is the first VNF placement scheme where an end-point of a service chain cannot be known prior to the placement.\\The second reason is that some of the previously reviewed works do not meet all our requirements. The ILPs proposed by \cite{Luizelli} and \cite{Moens} for instance are not suitable for large scale environments. References \cite{Xia}-\cite{Mechtri} do not consider QoS (i.e., service delay threshold) and sharing VNFs among the chains. References \cite{Luizelli}-\cite{Bouet} do not take a complete cost model into account. Unlike them, our work is appropriate for a large scale environment while considering all the above-mentioned points.

\section{VNF Placement Problem}
\subsection{Problem Description}
This paper focuses on the proactive placement of VNFs. The placement problem is formalized as follows: Content providers request the CDN provider to deploy a set of value-added services. The request specifies the service-related parameters (e.g., the description of functionalities that constitute the services) including the QoS threshold to be satisfied.
For ease of reading, we specifically denote the surrogate servers containing content as content servers. Therefore,
given a content X, the content servers containing X, a set of end-users requesting the content, their workload and a set of services to be accessed by the end-users, 
the VNF placement and chaining problem consist of : (i) jointly finding the number and location of surrogate servers to host the VNF instances and the content servers having content X, (ii) chaining the VNFs and connecting end-users, while minimizing the cost of CDN provider and satisfying the QoS threshold of all services offered to the end-users.  

The reader should note that for each end-user, the delay for viewing the video is the sum of the following delays:
\begin{itemize}
    \item the delay from the content server (selected to stream the video) to the surrogate server hosting the first VNF of the chain,
      \item the delay for transmitting and processing the video in the chain, and
        \item the delay from the surrogate server hosting the last VNF of the chain to the end-user.
\end{itemize}

As the first delay is unknown when the VNFs are placed, the total delay might violate QoS requirements when end-users access value added services. In this paper, we jointly optimize the selection of serving surrogate servers and VNF placement. It is therefore assumed that the serving surrogate servers are selected by new functional entity the “VAS serving surrogate server selector” instead of the traditional CDN controller. The impact on CDN functioning remain minimal since this new entity (which incorporates the results of our novel joint optimization algorithm) might be incorporated in the traditional CDN controller. 
 Our problem is a slight variation of the well-known Bin Packing problem \cite{Monaci} and the Hierarchical Facility Location-Allocation problem \cite{Teixeira200892}\cite{Farahani}.
Our problem can be an extension of the Bin Packing problem to Bin Packing plus Packer problem, including selecting the bins (i.e., surrogate servers) to drop the items (i.e., VNFs). The capacity of the bins and the load (i.e., end-user workload for a given VNF) of the items are known. The goal is to optimize the sum of the cost of the bins and the communication cost between the packers (i.e., end-users) and the bins.
In the Hierarchical Facility Location-Allocation problem, the sites are the equivalent of surrogate servers and each demand location is equivalent to end-user locations in our case. The facilities to be placed are equivalent to the VNFs in our case. The problem is decomposed into two phases: The location phase and the allocation phase, as follows: The location phase consists of determining the number of required facilities and the site to locate them. The allocation phase is for allocating the corresponding demand to each facility in the last layer (i.e., the last VNF of the VNF chain in our case). Besides, in the allocation phase, each facility (i.e., VNF) in each layer is allocated to one of the facilities (i.e., VNF) in the previous layer, which is equivalent to placing the instances of a VNF chain by connecting surrogate servers. The main difference between the Hierarchical Facility Location-Allocation problem and our VNF placement problem is that, in the former problem, the location of the primary level facilities (i.e., the first VNFs of the VNF chains in our case) are known beforehand, whereas, in our VNF placement problem, it is not.

\subsection{ILP Formulation}

Let us consider $N$ as a set of surrogate servers, $W$ as a set of content servers and $U$ as a set of end-users. The physical topology of the network is represented by a directed graph $G=(V, E)$, where $V=N \cup W \cup U$ is the set of nodes composed of the surrogate servers, content servers and end-users connected by directional edges $E$. 
Let us also consider $K$ as a set of VNFs of different types, such as Video Transcoder, Video Mixer and Video Compressor (c.f. Section 2.1). Each VNF of type $k \in K$ has a predetermined resource requirement and processing capacity, $R_k$ and $P_k$, respectively. The VNF requirement ($R_{k}$) can be a vector of vCPU, memory, disk, and I/O bandwidth (i.e., VNF processing capacity). If we assume that the bandwidth of the links connected to the NFVI node is less than the input/output bandwidth of that NFVI node, considering the I/O bandwidth in the requirement vector of VNFs avoids link overload.
The CDN provider also delineates a set of instances for each VNF type $k \in K$, $I_k$. Such specification may be the result of the license model adopted while acquiring the VNFs from the VNF provider or as a result of management restrictions.

A service request $h_{f}\in H$, generated by an end-user $f\in U$, is represented by a directed graph $\mathcal{G}(\mathcal{V}^{f}, \mathcal{E}^{f})$, where $\mathcal{V}^{f}$ is the set of VNFs that will be installed on nodes in $N$ and $\mathcal{E}^{f}$ is the set of virtual edges. 
For the sake of simplicity, we further define the following sets defined by:\\

\begin{equation*}
\begin{array}{lll}
&\Lambda^f : \{a~|~a~\text{is first element of VNF chain}~h^f \}\\
& \Delta^f : \{z~|~z~\text{is last element of VNF chain}~h^f \}\\
&\Phi^f : \{n \in W ~|~w_n^f = 1\}
\end{array}
\end{equation*}\\
where we denote respectively $\Lambda^f$ as the singleton containing the first VNF within the VNF chain $h_f$, $\Delta^f$ as the singleton containing the last VNF within the VNF chain and $\Phi^f$ as the set containing content servers which host the requested content for the service request $h_f \in H$.\\
Table \ref{variables} delineates the inputs and variables used in our ILP formulation.
{
\renewcommand{\arraystretch}{1.5}
\begin{table}[h]
\centering
\caption{Input Parameters and Variables}\label{variables}
\label{my-label}
\resizebox{
\columnwidth}{!}{%
\begin{tabular}{|p{1cm}|p{7cm}|}
\hline
\rowcolor{lightgray}\multicolumn{2}{|c|}{\textbf{Network Inputs}}                                                                                                                                         \\ \hline
$N$                 & Set of surrogate servers in the network, $N\subseteq V$                                                                                                          \\ \hline
$U$                 & Set of end-users in the network, $U\subseteq V$                                                                                                                  \\ \hline
$W$                 & Set of content servers in the network, $W\subseteq V$                                                                                                                  \\ \hline
$E$                 &The set of edges (i.e., logical communication links) in the network                          \\ \hline
$BW_{(u,v)}$        & The bandwidth capacity of edge $(u, v)\in E$                                                                                                          \\ \hline
$D_{(u,v)}$         & Delay of unit load (1 Gbps) for edge $(u, v)\in E$                                                                                                     \\ \hline
$\sigma(u,v)$       & Hop count of the edge $(u, v)\in E$                                                                                                                    \\ \hline
$B_n$               & Bandwidth cost (in dollars) of unit load (1 Gbps) per hop from surrogate server $n$                                                                                     \\ \hline
$\beta^f_{(u,v)}$    & Bandwidth cost incurred by sending load of end-user $f$ along edge $(u, v)\in E$                                                                         \\ \hline
$C_n$               & Capacity of surrogate server $n \in N$ in terms of resource units                                                                            \\ \hline
$\gamma$            & Site license cost                                                                                                                                                \\ \hline
$\delta_{n}$        & Operational cost for unit resource (vCPU) for surrogate server $n$                                                                                         \\ \hline
\rowcolor{lightgray}\multicolumn{2}{|c|}{\textbf{Service Inputs}}                                                                                                                                          \\ \hline
$H$                 & Set of services $h_f$ requested by user $f\in U$ \\ \hline
$K$                 & Set of VNF types that constitute all services $h_{f}\in H$                                                                                                       \\ \hline
$\mathcal{V}^{f}$              & $\mathcal{V}_{f}\subseteq K$ is a set of VNFs that constitute service $h_{f}\in H, f\in U$                                                                                 \\ \hline
$\mathcal{E}^f$              & Set of VNF edges of the VNF chain for service $h_{f}\in H$ , $f\in U$                                        \\ \hline
$\Delta^f$               &  The first VNF in the VNF chain for service $h_{f}\in H$, $f\in U$                                      \\ \hline
$\Lambda^f$               &  The last VNF in the VNF chain for service $h_{f}\in H$, $f\in U$                                       \\ \hline
$\Phi^f$               & Set of content servers hosting content requested for service $h_{f}\in H$, $f\in U$                                       \\ \hline
$I_{k}$             & Set of VNF instances of type $k \in K$                                                                                                                           \\ \hline
$\alpha_{k}$        & Software license cost of a VNF instance of type $k \in K$                                                                                                        \\ \hline
$T_{k,n}$           & Processing delay of VNF instance of type $k \in K$ on surrogate server $n\in N$ for unit load (1 Gbps)                                                           \\ \hline
$R_{k}$             & Resource requirement for VNF type $k \in K$                                                                                                     \\ \hline
$P_{k}$             & Processing capacity (Gbps) of VNF type $k \in K$                                                                                                                 \\ \hline

$L_{f}$             & Load of end-user $f \in U$                                                                                                                                       \\ \hline
$D_{{h}_{f}}$       & QoS (i.e., Service Delay),threshold of service $h_{f}\in H$                                                                                                       \\ \hline
\rowcolor{lightgray}\multicolumn{2}{|c|}{\textbf{Variables}}                                                                                                                                               \\ \hline
$x_{k,n,j}$         & 1, if instance $j$ of VNF type $k$ is assigned to surrogate server $n \in N$ and 0, otherwise                                                                      \\ \hline
$\lambda^{f}_{k,n,j}$ & 1, if VNF type $k$ belonging to VNF chain of  end-user $f$ is mapped to its instance $j$ on surrogate server $n$ and 0, otherwise                                    \\ \hline
$y_{u,v}^{f,p,q}$   & 1, if edge $(u,v)$ hosts VNF edge $(p, q)$ of VNF chain of  end-user $f$ and 0, otherwise                                                                          \\ \hline
$z_n$               & 1, if surrogate server $n$ is used and  0, otherwise \\                                                                                        \hline 
$w^f_n$             & 1, if a content server $n$ has the content requested for service $h_{f}\in H$ and  0, otherwise \\                                                                                        \hline 
\end{tabular}
}
\end{table}
}

\begin{figure*}[t]
\textbf{\textit{Objective function:}}\\
\begin{equation}\label{objective}
min\left (\sum_{\substack{\forall k \in K\\
                  \forall n \in N\\
                  \forall  j \in I_{k}}}
        x_{k,n,j}\cdot {\alpha_{k}} + 
          \sum_{\substack{\forall f \in U\\
                  \forall (p,q) \in \mathcal{E}^f\\
                  \forall (u,v) \in E}}
        \beta_{(u,v)}^f \cdot y_{u,v}^{f,p,q}+
        \sum_{\substack{\forall f \in U\\
                  \forall  k \in \Lambda^f\\
                  \forall w \in \Phi^f\\
                \forall n \in N\\
                 \forall j \in I_k\\
                \forall (w,n) \in E}}
\beta_{(w,n)}^f \cdot \lambda^{f}_{k,n,j} +
 \sum_{\substack{\forall f \in U\\
                  \forall  k \in \Delta^f\\
                \forall n \in N\\
                 \forall j \in I_k\\
                \forall (n,f) \in E}}
\beta_{(n,f)}^f \cdot \lambda^{f}_{k,n,j} +
(z_n \cdot \sum_{\forall n \in N}(\gamma + \sum_{\substack{ \forall n \in N\\
 \forall k \in K\\
 \forall  j \in I_k }} 
 R_k \cdot \delta_n \cdot x_{k,n,j}))\right)
\end{equation}
\end{figure*}

\begin{figure*}[!h]
\textbf{\textit{QoS guarantee:}}\\
\begin{equation}\label{cst7}
\sum_{\substack{\forall (p,q)\in \mathcal{E}^f\\ \forall (u,v)\in E\\ \forall f\in U}}D_{(u,v)}\cdot L_f\cdot y_{u,v}^{f,p,q}+
\sum_{\substack{\forall k\in \mathcal{V}^f \mid \\ \forall i\in \mathcal{V}^f \\ \exists (k,i)\in E_f\\ \forall  u\in N \\ \forall  j\in I_k \\ \forall f\in U}} T_{k,u}\cdot L_f\cdot \lambda^f_{k,u,j}+
\sum_{\substack{\forall f \in U \\\forall  k \in \Lambda^f \\ \forall w \in \Phi^f \\ \forall n \in N \\\forall  j \in I_k \\ \forall(w,n) \in E}}(T_{k,n} + D_{w,n}) \cdot L_f \cdot \lambda^{f}_{k,n,j} + 
\sum_{\substack{\forall f \in U \\ \forall  k \in \Delta^f \\ \forall n \in N \\ \forall  j \in I_k \\ \forall (n,f) \in E}}{ (T_{k,n} + D_{n,f}) \cdot L_f \cdot \lambda^{f}_{k,n,j} }\preceq D_{{h}_{f}}
\end{equation}
\end{figure*}

Our objective is to minimize the cost of VNF placement for value-added services in CDN, including the cost of deploying VNF instances, the cost of using surrogate servers and the cost of communication as shown in (\ref{objective}). The cost of deploying VNFs is characterized by a software license cost per instance denoted $\alpha_k$. The cost of using a surrogate server $n \in N$ is the sum of a fixed license cost $\gamma$ and the operational costs for all VNF type instances. The license cost is the same for every surrogate server. The operational cost for a VNF instance of type $k$ is $R_k\cdot \delta_n$. The cost of communication in VNF placement for value-added services is the sum of the bandwidth costs amongst each pair of surrogate servers $u,v \in N$ hosting VNFs of the VNF chain in each end-user service request $h_f\in H$,$f \in U$, $\beta_{(u,v)}^f$. 
It further includes the bandwidth costs $\beta_{(n,f)}^f$ between the surrogate servers $n\in N$ hosting the tail VNFs of the VNF chain to an end-user $f \in U$ and the bandwidth costs $\beta_{(w,n)}^f$ between the surrogates servers $n \in N$ hosting the head VNFs of the VNF chain to a content server $W \in \Phi^f$ hosting the content requested by the user $f \in U$. Note that we compute the cost of using bandwidth $\beta_{(u,v)}^f$ between $u\in V$ and $v\in V$ using $\beta_{(u,v)}^f = L_f\cdot \sigma_{(u,v)}\cdot B_u$ with hop count $\sigma_{(u,v)}$, for load $L_f$ for end-user $f \in U$.
\\

\textbf{\textit{Content server selection:}}

\begin{equation}\label{serverSelection}
    \sum_{n\in W | n \in \Phi^f} w^f_n = 1,  \quad \forall  h_f\in H
\end{equation}

\textbf{\textit{VNF Placement:\\}}
\begin{equation}\label{cst1}
    \sum_{n\in N} \sum_{j\in I_k}\lambda^f_{k,n,j} \geq 1,  \quad \forall  f\in U,k\in K
\end{equation}

\begin{equation}\label{cst2}
\sum_{f\in U}L_{f}\cdot\lambda^f_{k,n,j}\preceq P_{k}\cdot x_{k,n,j}  \quad \forall k\in K, n\in N, j\in I_k
\end{equation}

\begin{equation}\label{cst3}
\sum_{k\in K}\sum_{ j\in I_k}R_k\cdot x_{k,n,j}\preceq C_n\cdot z_n  \quad \forall n\in N
\end{equation}
\\
\textbf{\textit{VNFs chain mapping:\\}}

\begin{equation}\label{cst4}
\begin{aligned}
\sum_{j\in I_p}\lambda^f_{p,u,j}\cdot & \sum_{j'\in I_q}\lambda^f_{q,v,j'}=y_{u,v}^{f,p,q} \\
\quad & \forall f\in U, (p,q)\in \mathcal{E}^f, (u,v)\in E
\end{aligned}
\end{equation}

\begin{equation}\label{cst5}
\sum_{(u,v)\in E}y_{u,v}^{f,p,q}=1 \quad \forall f\in U, (p,q)\in \mathcal{E}^f
\end{equation}

\begin{equation}\label{cst6}
\sum_{f\in U}\sum_{(p,q)\in \mathcal{E}^f} L_f\cdot y_{u,v}^{f,p,q}\preceq BW_{(u,v)} \quad \forall(u,v)\in E
\end{equation}
\\
Variables $x_{k,n,j}$ are used to identify unique instance $j\in I_k$ of VNF type $k\in K$ installed on surrogate server $n\in N$. Variables~$z_n$ are used to record surrogate servers hosting VNFs.

Constraint (\ref{cst7}) is the QoS constraint that guarantees that the end-user service request is delivered within the predefined delay threshold. The delay in delivering the service consists of two components, the network communication delay and the VNF processing delay on surrogate servers.
The first term of constraint (\ref{cst7}) calculates the network communication delay as the sum of delay between each pair of surrogate servers hosting the VNFs of the VNF chain in the end-user service requests.  The second term of constraint (\ref{cst7}) denotes the processing delay of VNFs hosted surrogates servers, which is proportional to end-user load assigned to the VNFs.
In the third term, we consider the communication delay $D_{w,n}$ between the content server $w \in \Phi^f$, selected to serve the service request and the surrogate server $n \in N$ hosting the head VNF of the VNF chain. We further include the variable $T_{k,n}$ to denote the processing delay incurred at the surrogate server $n \in N$. Similarly, the fourth term represents the  communication delay $D_{n,f}$ between the surrogate servers $n\in N$ hosting tail VNFs to the end-users $f \in U $. Naturally, we also take into account the processing delay $T_{k,n}$ incurred at the servers $n$ during the transmission to the end-users $f$.
Hence, constraint (\ref{cst7}) represent the end-to-end service delay from content server selection until end-user delivery.

We also ensure through constraint (\ref{serverSelection}) that only one content server $n \in W$ is selected to serve the request from end-user $f \in U$.

We guarantee in (\ref{cst1}) that, for each end-user service request $h_f\in H$, $f\in U$, an instance of the requested VNF type must be assigned to a surrogate server. It is assumed that an instance of a VNF type accommodates the load from an end-user. Though an instance of a VNF type can cater to multiple users, it must satisfy at least one end-user load. This can be trivially extended to split larger loads of an end-user into smaller chunks, since our formulation allows end-users $f$ and $f'$ to request the same service, that is $h_f=h_{f'}$.

Constraint (\ref{cst2}) ensures that the capacity of an instance of a VNF of type $k \in K$ is not exceeded by the total load requested by all the end-users assigned to it while constraint (\ref{cst3}) ensures that the total resource required by instances of all VNF types does not exceed the capacity of the host surrogate server.

In our model, we map the nodes (i.e., the VNFs) and their edges (i.e., the chain), in each end-user service request to the physical network in $G$. Therefore, the edge $(p,q)$ between two consecutive VNFs in each end-user service request must be assigned to a physical edge $(u,v)$ between two surrogate servers $u$ and $v$, in (\ref{cst4}). It should be noted that (\ref{cst4}) is a non-linear constraint and can be trivially linearized by replacing it with linear constraints (\ref{cst8})-((\ref{cst10})) as follows. \\

\begin{equation}
\begin{aligned}
\label{cst8}
y_{u,v}^{f,p,q}\preceq \sum_{j\in I_k} \lambda^f_{p,u,j} \\
\quad \forall f\in U,(p,q)\in \mathcal{E}^f , (u,v)\in E, u\in N , v\in N
\end{aligned}
\end{equation}

\begin{equation}
\begin{aligned}
\label{cst9}
y_{u,v}^{f,p,q}\preceq \sum_{j\in I_k} \lambda^f_{q,v,j} \\
\quad \forall f\in U,(p,q)\in \mathcal{E}^f , (u,v)\in E, u\in N , v\in N
\end{aligned}
\end{equation}

\begin{equation}
\begin{aligned}
\label{cst10}
y_{u,v}^{f,p,q}\succeq \sum_{j\in I_k} \lambda^f_{q,v,j}+\sum_{j\in I_k} \lambda^f_{p,u,j}-1 \\
\quad \forall f\in U,(p,q)\in \mathcal{E}^f , (u,v)\in E, u\in N , v\in N 
\end{aligned}
\end{equation}

As is ensured in (\ref{cst5}), the VNFs and their respective ordered edges are mapped to only one pair of physical surrogate servers and their edge. With respect to the underlying physical network, we guarantee that the total load on an edge in the physical network does not exceed the bandwidth capacity, in (\ref{cst6}).

The VNF placement and chaining for value-added services in CDN is an NP-Hard problem, calling for an efficient heuristic.

\section{Cost-efficient Proactive VNF Placement (CPVNF)}\label{PCPV}
In this section, we discuss the design choices and insights for our Cost-efficient Proactive VNF Placement heuristic for deploying value-added services in CDN.
There are three main constraints for VNF placement problem that should be considered: satisfying QoS as well as preventing VNF and server overloading.\\ We base our heuristic on the PageRank algorithm \cite{BRIN1998107} pioneered on the Google search engine. It can be described as a variant of the eigenvector centrality method. PageRank has also been known to perform well in scale-free networks and thus has been widely employed in many fields \cite{Cheng2011}. In summary, page ranking leverages the idea that a web page's importance is a factor of both the quantity and the quality of the pages linked to it. Comparatively, a surrogate server's fitness to host a VNF instance can also be viewed as dependent of the quality of the surrogate server itself (e.g capacity, processing power, energy consumption, etc.) as well as the quality and quantity of outgoing links towards other VNF instances in the VNF chain.  

We start by ordering every user request received at a given time period $t$ based on its aggregated requirements (e.g. vcpu, processing capacity,etc.). This is done, in supposition that user requests with stricter QoS requirements may be harder to map at later stages in the network. In a sense that, as we map each request, the remaining surrogate capacity and edge bandwidth capacity lessens within the network, thereby rendering subsequent requests mapping more complex.

\subsection{Selecting the content server}\label{findContentServer}
As previously mentioned in subsection \ref{illustrationCase}, a given content is duplicated over several content servers as per CDN principles. Hence, we must determine, upon choosing a surrogate server to host the first VNF in the VNF chain, the content server upon which the content is routed from. The choice of a content server to provide for a request is an important factor towards guaranteeing QoS requirements in terms of delay. Therefore, it also influences the fitness of the surrogate server hosting the first VNF. Hence, we assume there is a convex function $Q: \mathbb{R} \rightarrow \mathbb{R}_+$, which computes the fitness of a content server from a surrogate server perspective and also acts as a penalty function penalizing content servers with large delays to the surrogate server hosting the first VNF. We hence define $Q(\cdot)$ as a quadratic function, widely used in the control theory literature such that 
\begin{equation}\label{quad}
Q(w,n) = \mu\cdot(\frac{D(w,n)}{D_{h_f}})^2
\end{equation}
with $w \in \Phi^f$, $(w,n)$ denoting the path with the least delay between $w \in \Phi^f$ and $n \in N$. Where $D(w,n)$ represents the delay between one of the content servers $w$ and a surrogate server $n \in N$ candidate towards hosting the first VNF of the chain, $D_{h_f}$ is the delay threshold for user request $h_f$ and $\mu$ is a constant.
Note that in this context, the penalty function obviously penalizes content servers which predominantly contributes to the violation of the delay constraint (detailed below).
\subsection{Selecting the surrogate servers}
As evident from our objectives previously stated, we are motivated to guarantee SLA bounds on the QoS, with respect to end-user perceived latency as well as preventing VNF and server overloading.
In this regard, it is critical for the VNF mapping process to place VNFs onto surrogate nodes which contribute most to this goal. We therefore employ a surrogate importance rank metric (SIR), we note $\varphi_{k,n}$ to infer the relative fitness of a given candidate surrogate server $n$ to host a VNF instance $k$ in comparison to its peers. $\varphi_{k,n}$ indicates the surrogate server importance rank with regard to the current network topology properties such as the available bandwidth between edges, the remaining capacities of reachable surrogate servers, etc.
A surrogate server's SIR is determined notably by its remaining capacity, the aggregated remaining bandwidth of its outgoing links, the rank of its neighbours as well as a weight $\Gamma$ to denote whether the nodes already hosts an instance of the VNF considered. The latter parameter is to help ensure that a minimal number of instances are used throughout the network in order to reduce costs. We justify considering the number of outgoing links in our model as it also increases the probability to reach the requesting end-users and thus ensure that the location of the last VNF is accessible to the requesting end-user. Of course, all of this is done while always keeping in mind to respect QoS constraints. The actual value of the instance weight can be determined experimentally by finding a fair tradeoff between VNF instance consolidation, server license costs and success of QoS guarantee. We therefore consider
\\
\begin{equation}
 M_n = (\pi \cdot C'_n) \cdot [(1-\pi) \sum_{a \in A(n)} BW'(n,a)] \quad \forall n\in N
\end{equation}
\\
where $C'_n$ and $BW'(n,a)$ respectively denote the remaining capacity at surrogate server $n$ and the remaining bandwidth capacity between edge $(n,a)$. The set $A(n)$ representing the surrogate servers adjacent to the surrogate server $n$. The weight $\pi$ is used to bias node selection by focus on surrogate servers' capacity or aggregate outgoing bandwidth  depending on network topology state. Indeed, if we suppose a scenario where a majority of surrogates servers have high capacity, the VNF mapping should focus more on the quality of outgoing links towards other surrogates servers and vice-versa. Hence, these parameters can be seen as a intensification or diversification parameter during the search process of the most suitable candidate surrogate servers.

Upon launching our heuristic and before any user request is mapped, the initial values of each surrogate server's SIR is determined by : 
\begin{equation}\label{initialSIR}
 \varphi_{k,n} = \frac{M_n}{\sum_{\substack{o\in N}} M_o } \quad \forall n \in N
\end{equation}

Once initialized, we make use of personalized pagerank where we bias towards nodes with preexisting instance of the vnf type $k$ to reduce the number of surrogate servers used :
\begin{equation}\label{SIRcomp}
   \varphi_{k,n} = \Gamma_k * \frac{1-\psi}{|N|} + \psi \cdot (\sum_{i \in A(n)  } \frac{\varphi_{k,i}}{|A(i)|} )
\end{equation}
with :
\begin{equation*}
  \Gamma_k =\begin{cases} \Gamma ~\text{if } x_{k,n,j} = 1\\ 
  1~ \text{otherwise} 
  \end{cases}
\end{equation*}

where $\psi$ denotes a damping factor usually set optimally to 0.85 \cite{Son2012} and $|A(i)|$ the amount of neighbouring surrogates servers to server $i$.
Note that the SIR values of each surrogate server can be effectively computed using both iterative and algebraic methods \cite{Page1998}.
Once the SIR values computed, we proceed towards mapping the VNF chain in two phases : the VNF to surrogate server mapping and the VNF to VNF link mapping.

\begin{algorithm}[h]
\caption{CPVNF}\label{alg:CPVNF}
\DontPrintSemicolon

\KwIn{Network topology, $H$:user requests, $N$ : surrogate servers, $stop$ : search limit }
\KwOut{VNF mapping}
\SetKwBlock{Begin}{function}{end function}
     H' = \{\}\;
     searchLimit = 0\;
     \For{$n \in N$}{
        compute initial SIR values using eq. \ref{initialSIR}
     }
    \For{$h_f \in H$}{
        H'= rank ($h_f \in H$) based on added requirements 
    }
    \For{$h'_f \in H'$}{
        \For{$k\in h'_f$}{
            \eIf{k : first VNF of the chain}{
                \For{$n \in N$}{
                    compute SIR of $n$ for VNF $k$ using eq. \ref{initialSIR}-\ref{SIRcomp}\;
                    compute Q($\cdot$) for every content server $w$ using eq.\ref{quad} 
                }
                retain a host for VNF $k$, surrogate server $n$ with highest compound score ($SIR+\frac{1}{Q(w,n)}$)
            }{
                \For{$n \in N$}{
                    compute SIR of $n$ for VNF $k$ using eq. \ref{SIRcomp}\;
                    rank $n$ based on SIR value\;
                    
                }
                \eIf{path between n and surrogate hosting previous vnf in chain}{
                    place VNF $k$ on highest ranked $n$
                    
                }{
                    select next ranked $n$ 
                }
            }
            \For{$k,k' \in h'_f$}{
                find k-shortest paths of hosts of $k$ and $k'$\;
                select path with lowest delay
            }
            \For{k:last VNF in the chain}{
                find k-shortest paths between $k$ to end-user $f$ \;
                select path with lowest delay\;
            }
        }
        \eIf{QoS constraints respected}{
            restore $\pi$ and $\kappa$ to default values\;
            confirm removal of server and edges capacities
            continue to next request\;
        }{
            \eIf{searchLimit < stop}{
                decrease $\pi$
                redo for loop above\;
            }
            {
            reject request\;
            }
        }
    }
    
\end{algorithm}

In the VNF to surrogate mapping, we first sort each surrogate server according to their SIR to determine their fitness. We attempt to place the VNFs of the VNF chain accordingly. However, keep in mind that we must take into account the selection of content server for the first VNF. Hence, the selection of the surrogate server to host the first VNF in the VNF chain is done based not only on its SIR value but also on the penalty incurred by the content source as previously described in subsection \ref{findContentServer}. Otherwise, a VNF is placed on the best ranked surrogate server.
After a VNF is placed on a given surrogate server, the SIR values of every surrogate server in the network is updated to reflect the current network topology state.
The VNF to VNF link mapping consists of finding the k-shortest paths between the surrogates servers hosting VNFs for the VNF chain.
A given path is retained when it satisfies the specified delay requirements. Note that the delay between the last VNF of the chain and the requesting end-user is also included for the path choice.
We summarize the VNF mapping process described above in Algorithm \ref{alg:CPVNF}.

\section{performance Evaluation}
This section describes our simulation settings and presents the results of the performance evaluation of our proposed proactive VNF placement solution in CDN networks. As mentioned earlier, this is the first contribution on VNF chaining placement in CDN where the service chain format boundary (i.e., source and destination of the service chain) is different. Thus, CPVNF cannot be compared with existing state-of-the-art algorithms. Besides, as basic algorithms such as random and greedy VNF placements usually do not consider QoS (e.g., service threshold delay), therefore comparing CPVNF to them is not adequate. 
To evaluate the effectiveness of our proposed algorithm, we compare its performance against the optimal solution of the ILP model for a relatively small environment size. Next, extensive simulations, using several setting parameters, are driven to further evaluate our algorithm. 

{\renewcommand{\arraystretch}{1.5}
\begin{table}[h]
\centering
\caption{Simulation Parameters \cite{Bouet,Cacheda2007}}
\label{Simulation Parameters}
\resizebox{\columnwidth}{!}{%
\begin{tabular}{c|c|c|c|}
\cline{2-4}
\multicolumn{4}{|c|}{\textbf{Parameters}}\\ \hline
\multicolumn{2}{|c|}{Number of servers}  &  \multicolumn{2}{c|}{9}\\ \hline
\multicolumn{2}{|c|}{Number of end-users}&\multicolumn{2}{c|}{$9,12,15,18,25$}   \\ \hline
\multicolumn{2}{|c|}{Number of VNFs per chain} & \multicolumn{2}{c|}{3}\\ \hline
\multicolumn{2}{|c|}{Service Delay Threshold (ms) ($D_h^{th}$ in ILP) }& \multicolumn{2}{c|}{80-250} \\ \hline
\multicolumn{2}{|c|}{Bandwidth Cost (Dollar/Gbps) (B in ILP)}  &  \multicolumn{2}{c|}{10} \\ \hline
\multicolumn{2}{|c|}{End-user Load (Mbps) ($ L_f$ in ILP)}   & \multicolumn{2}{c|}{15-50} \\ \hline
\multicolumn{2}{|c|}{Surrogate servers capacity (vCPU) $C_n$ in ILP)} & \multicolumn{2}{c|}{16-64} \\ \hline
\multicolumn{2}{|c|}{Site license cost(Dollar)($\gamma$ in ILP)} & \multicolumn{2}{c|}{1000} \\ \hline
\multicolumn{2}{|c|}{\begin{tabular}[c]{@{}c@{}}Surrogate servers Operational cost (Dollar/vCPU)\\ ($\delta_n$ in ILP)\end{tabular}}   & \multicolumn{2}{c|}{5-10}          \\ \hline
\multicolumn{2}{|c|}{\begin{tabular}[c]{@{}c@{}}VNF license cost (Dollar/vCPU)\\ ($\alpha_{k}$  in ILP)\end{tabular}}  & \multicolumn{2}{c|}{100}                                                    \\ \hline
\multicolumn{2}{|c|}{\begin{tabular}[c]{@{}c@{}}Capacity weight ($\pi$)\\ \end{tabular}}  & \multicolumn{2}{c|}{0.8}                                                    \\ \hline
\multicolumn{2}{|c|}{\begin{tabular}[c]{@{}c@{}}VNF instance weight ($\Gamma$)\\ \end{tabular}}  & \multicolumn{2}{c|}{2}                                                    \\ \hline
\multicolumn{2}{|c|}{\begin{tabular}[c]{@{}c@{}}Penalty function coefficient  ($\mu$)\\ \end{tabular}}  & \multicolumn{2}{c|}{0.5}                                                    \\ \hline
\end{tabular}}
\end{table}}

\subsection{Experimental Set up}

We setup our network topologies using a customized version of the topology generator tool : topology-generator \cite{Cesar2015}. More specifically, we generate multiple scenarios with the objective of evaluating the influence of several end-user, network, surrogate server and VNF type related parameters with regard to the VNF placement problem. Hence, while retaining the core network topology (i.e. edge connections between surrogates), we vary characteristics such as server capacity, edge bandwidth, VNF resource requirements etc.
In summary, we consider a base scenario in which 9 surrogate servers are randomly interconnected each by 1 to 4 outgoing links towards other surrogate servers. Furthermore, we consider in our simulations, 5 content servers uniformly spread across the network, from which content needed to provide for service requests are stored. For added realism, we ensure that each content can be found in at least 3 of the 5 content servers as expected in CDN networks where content are duplicated. Furthermore, We configure each content server to randomly have 1 to 3 outgoing links towards surrogates servers in the network. Finally, we vary the number of end users. It is important to note that we consider directed edges between nodes and the bandwidth capacity of each edge is randomly taken between 100, 1000 and 10000 Mbps. Moreover, we consider for each service request, a delay threshold ranging from 80 to 250(ms) based on reference \cite{Cacheda2007}.
All simulation and network topology parameters are summarized respectively in Table \ref{Simulation Parameters} and \ref{topologyDescription} with VNF license cost and site license cost selected from \cite{Bouet}.

{\renewcommand{\arraystretch}{1.5}

\begin{table}[h]
\centering
\caption{Topology parameters}
\label{topologyDescription}

\begin{tabular}{|l|l|}
\hline
\textbf{Parameters} & \textbf{Values} \\ \hline
Number of surrogate servers & 9 \\ \hline
Number of end-users & 9,12,15,18,25    \\ \hline
Number of content servers & 5   \\ \hline
Bandwidth per edge (Mbps) & (100,1000,10000)  \\ \hline
Max \# of outgoing links &  1-4 \\ 
surrogate to surrogate &  \\  \hline
Max \# of outgoing links & 1-2   \\ 
end user to surrogate & \\ \hline
Max \# of outgoing links  & 1-3 \\ 
content server to surrogate  & \\ \hline
\end{tabular}

\end{table}
}

\subsection{Performance Metrics}
To evaluate the effectiveness of the proposed algorithm, four performance metrics are taken into account.
\begin{figure*}[h]
\centering
\begin{minipage}[b]{.4\textwidth}
 \includegraphics[width=\textwidth]{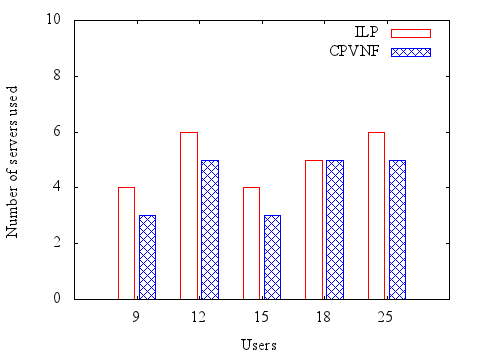}
    \caption{Number of servers used}
    \label{fig:num_servers}
\end{minipage}\qquad
\begin{minipage}[b]{.4\textwidth}
    \includegraphics [width=\textwidth]{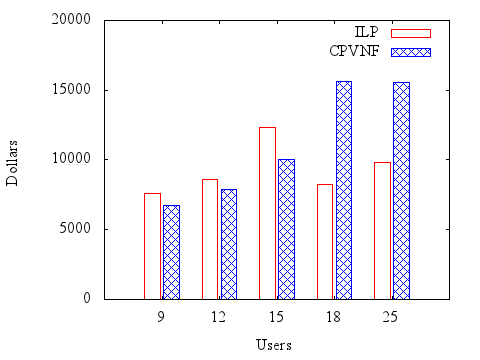}
    \caption{Operational cost}
    \label{fig:op_cost}
\end{minipage}

\end{figure*}
\begin{enumerate}
    \item \textit{Operational cost}: It is defined as the total cost of using the surrogate servers on which VNFs are deployed as well as the VNF and site license cost. It is expressed in dollars. 
     \item \textit{Communication cost}: It is defined as the total bandwidth cost incurred by serving the end-users service requests. It includes the cost of communication between the content server to the surrogate server hosting the first VNF, between the surrogate servers that host VNFs and the cost of communication between the surrogate servers that host the last partition (or last VNF) of the VNF chain and the end-user. 
    \item \textit{Total cost (Dollar)}: It is defined as the sum of operational cost and communication cost. Reduced total cost indicates the cost-efficiency of a VNF placement algorithm.
    \item \textit{Average response time :} It is defined as the average duration in which a given content is retrieved from a selected content server, traverse the required surrogate servers hosting the VNFs specified in its VNF chain to its final delivery to the end-user.
\end{enumerate}

\subsection{Results and Discussions}
\begin{figure*}[t]
\centering
\begin{minipage}[b]{.4\textwidth}
    \includegraphics[width=\textwidth]{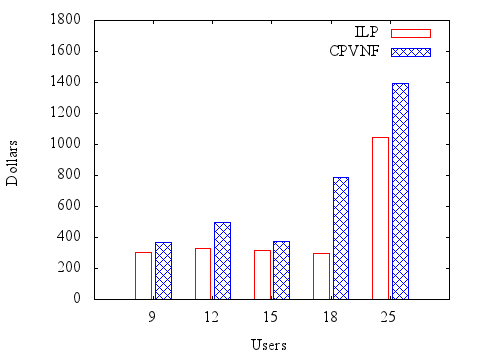}
    \caption{Communication cost}
    \label{fig:com_cost}
    
\end{minipage}
\qquad
\begin{minipage}[b]{.4\textwidth}
    \includegraphics [width=\textwidth]{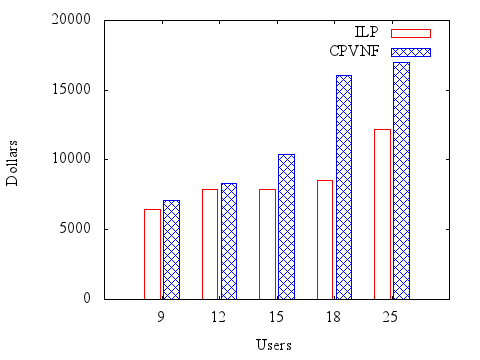}
    \caption{Total cost}
    \label{fig:total_cost}
\end{minipage}
\end{figure*}

Our heuristic CPVNF, heavily focuses on reducing the operational cost in order to achieve a reduction of the total cost. As such, we initially set the capacity weight $\pi$ at 0.8 such that surrogate servers with higher capacities are favoured. Furthermore, once a VNF instance is scheduled for a given surrogate server, we apply a VNF instance $\Gamma_k$ to ensure continued utilization of the surrogate server. Hence, as it can be observed in Fig. \ref{fig:num_servers}, CPVNF globally allows for a slightly fewer number of surrogate servers to be used to host VNFs. It is worth noting however, that one should not assume that CPVNF performs better than the ILP solution. In fact, our strategy while globally beneficial in terms of operational cost as shown in Fig \ref{fig:op_cost} also has an adverse effect in terms of communications costs shown in Fig. \ref{fig:com_cost} where the ILP performs much better. Such trade-off between the number of servers used and the communication cost  allows for the ILP to obtain reduced total costs as illustrated in Fig. \ref{fig:total_cost}.

In short, our approach for CPVNF consist of finding the most appropriate surrogates, determine the shortest paths to both the selected content server and the end-user with the aim of respecting the required QoS threshold. We base this strategy and our trade-off choice upon the a priori fact that the site license cost is much higher than the bandwidth cost. Therefore, to focus on reducing the number of servers used is computationally more efficient than exploring optimal network paths as we keep in mind that we are also designing for large scale networks. Note however that, such placement policy does not always lead to a successful mapping with respect to the QoS constraints due to the fact that both content servers and end-users may be located relatively far from the surrogate servers resulting in higher number of hops and consequently more bandwidth resource consumption.

\begin{figure*}[t]
\centering
\begin{minipage}[b]{.4\textwidth}
    \includegraphics [width=\textwidth]{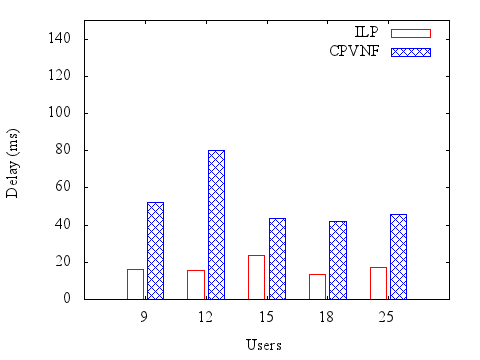}
    \caption{Average response time}
    \label{fig:response_time}
\end{minipage}\qquad
\begin{minipage}[b]{.4\textwidth}
    \includegraphics [width=\textwidth]{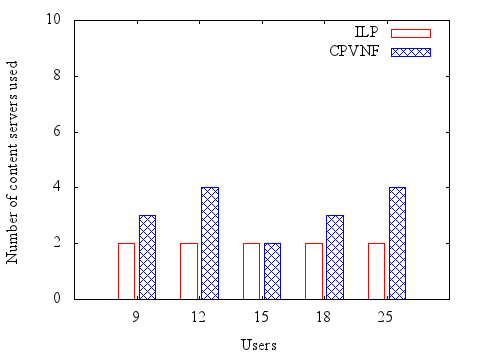}
    \caption{Number of content servers used}
    \label{fig:content_servers}
\end{minipage}\qquad
\end{figure*}

\begin{figure*}[t]
\centering
\begin{minipage}[b]{.4\textwidth}
    \includegraphics [width=\textwidth]{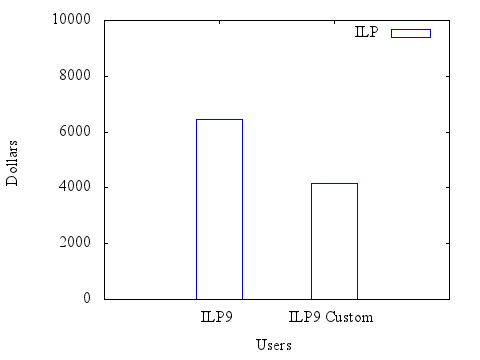}
    \caption{Comparison of total cost with reduced $\alpha_k$ and $\delta_n$ - 9 servers}
    \label{fig:total_cost2}
\end{minipage}\qquad
\begin{minipage}[b]{.4\textwidth}
    \includegraphics[width=\textwidth]{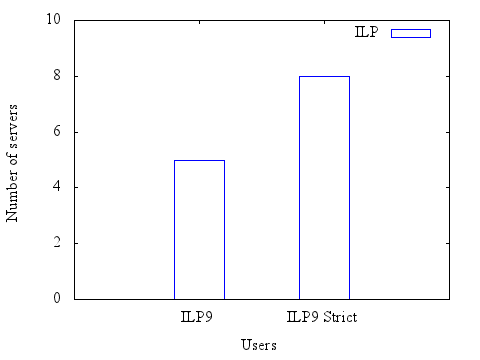}
    \caption{Number of server used with tighter specifications - 9 servers}
    \label{fig:num_servers2}
\end{minipage}

\end{figure*}

In this case, we allow several iterations in which we temporarily focus on nodes with higher outgoing bandwidth capacities rather than the node capacity itself by progressively reducing the node capacity weight $\pi$ in order to find more appropriate nodes. A consequence of this exploration is that we are using more content servers as shown in Fig. \ref{fig:content_servers} given the fact that our penalty function always favours the content servers with the least delay from a surrogate server's perspective. We do however argue that CPVNF could also achieve better results with an optimal value of $\pi$ which better reflect the network topology. Precise determination of such value is left for future work.
We further evaluated in Fig.\ref{fig:response_time} the average response time of requests in our simulations. While the ILP solution naturally achieves better response times for requests, those obtained by CPVNF are relatively low.
Another important characteristic of a VNF placement consists of its time complexity. It is shown that for a given precision parameter $\epsilon$, page rank can be iteratively computed with a number of iterations proportional to max\{1-$\epsilon$\}\cite{Bianchini2005, Cheng2011}. Furthermore, it is possible to solve the VNF to VNF link mapping problem in polynomial time according to \cite{Ahuja1993,Cheng2011}. Hence, CPVNF is a polynomial-time algorithm in terms of N, K and max\{1-$\epsilon$\}.

We are also interested in identifying which parameters has the most influence in VNF placement. Interestingly, in one scenario (15 users) illustrated in Fig. \ref{fig:num_servers}, we observe a fewer number of server used compared to scenarios with less end-users. This scenario is characterized by a large number of service requests composed of the same VNF chain combined to delay thresholds ranging from 127-250 (ms). We thus hypothesize that the characteristics of service requests (variety of VNF types across requests, delay threshold) have more influence than the number of end users. To confirm this, we evaluate in Fig. \ref{fig:num_servers2} the number of servers used as we tighten the delay threshold for service requests within the range of 40-90 (ms) and increase the resource requirement $R_k$ for VNF types to be randomly taken within the range of (4-8) while reducing surrogate servers capacity $C_n$ to be either 8 or 16. We then note, an increase in the number of servers used, obviously leading to higher operational costs.

In contrast, in our simulations, sensibly reducing the values of $\alpha_k$ and $\delta_n$ while maintaining vnf resource requirements and surrogate server capacity did not yield significant changes in vnf placement. Obviously, as shown in Fig. \ref{fig:total_cost2}, there is a reduction of the total cost mainly due to the multiplicative reduction of these parameters.
\section{Conclusion and future work}
This paper proposes a cost-efficient proactive VNF placement algorithm to guarantee the service delay in offering VNF based value-added services to end-users. The objective is to place VNFs in a way that it leads to the optimum number of VNFs to reduce the cost while still satisfying the service delay threshold. Providing a maximum delay to end-users located anywhere in the network is one of the main benefits of the proposed algorithm. An ILP has been presented to model the VNF placement problem for value-added services in CDNs. In addition, a systematic view of the VNF lifecycle has been presented. 
As the first to propose VNF chaining placement in CDNs, (since the service chain format of our work is different from those of related works), we only compare CPVNF with ILP for small size environments. 
As a future work, the proactive VNF placement algorithm will be integrated into an appropriate scaling algorithm to handle fluctuations in end-user workload over time. The scaling algorithm will be triggered in the case of violation in some performance indicators such as QoS and/or resource utilization at the run-time. Adding a prediction model to predict the future end-user workload in the environment and deploying VNF instances accordingly is also considered as one of the future extension of this paper.

\section*{Acknowledgment}
{This work was supported in part by Ericsson Canada, and the Natural Science and Engineering Council of Canada (NSERC). We would also like to thank Elaheh T. Jahromi for her contributions to the use case described in the paper.}
\bibliographystyle{IEEEtran}

\bibliography{TNSM}

\begin{thebibliography}{10}
\providecommand{\url}[1]{#1}
\csname url@samestyle\endcsname
\providecommand{\newblock}{\relax}
\providecommand{\bibinfo}[2]{#2}
\providecommand{\BIBentrySTDinterwordspacing}{\spaceskip=0pt\relax}
\providecommand{\BIBentryALTinterwordstretchfactor}{4}
\providecommand{\BIBentryALTinterwordspacing}{\spaceskip=\fontdimen2\font plus
\BIBentryALTinterwordstretchfactor\fontdimen3\font minus
  \fontdimen4\font\relax}
\providecommand{\BIBforeignlanguage}[2]{{%
\expandafter\ifx\csname l@#1\endcsname\relax
\typeout{** WARNING: IEEEtran.bst: No hyphenation pattern has been}%
\typeout{** loaded for the language `#1'. Using the pattern for}%
\typeout{** the default language instead.}%
\else
\language=\csname l@#1\endcsname
\fi
#2}}
\providecommand{\BIBdecl}{\relax}
\BIBdecl

\bibitem{Pallis:2006}
G.~Pallis and A.~Vakali, ``Insight and perspectives for content delivery
  networks,'' \emph{Commun. ACM}, vol.~49, no.~1, pp. 101--106, January 2006.

\bibitem{Pathan2008}
M.~Pathan, R.~Buyya, and A.~Vakali, \emph{Content Delivery Networks: State of
  the Art, Insights, and Imperatives}.\hskip 1em plus 0.5em minus 0.4em\relax
  Berlin, Heidelberg: Springer Berlin Heidelberg, 2008, pp. 3--32.

\bibitem{Akamai}
\BIBentryALTinterwordspacing
Trefis. (2015) Akamai q2 earnings preview: Value-added services will drive
  growth, but currency headwinds likely played the minor spoilsport. [Online].
  Available: \url{http://www.trefis.com/}
\BIBentrySTDinterwordspacing

\bibitem{Pathan2014}
M.~Pathan, R.~K. Sitaraman, and D.~Robinson, \emph{Advanced Content Delivery,
  Streaming, and Cloud Services}.\hskip 1em plus 0.5em minus 0.4em\relax Wiley
  Publishing, 2014.

\bibitem{Index}
Cisco, ``Cisco visual networking index: Forecast and methodology,
  2014–2019,'' Cisco, Tech. Rep., 2015.

\bibitem{Han}
B.~Han, V.~Gopalakrishnan, L.~Ji, and S.~Lee, ``Network function
  virtualization: Challenges and opportunities for innovations,'' \emph{IEEE
  Communications Magazine}, vol.~53, no.~2, pp. 90--97, February 2015.

\bibitem{Mijumbi}
R.~Mijumbi, J.~Serrat, J.~L. Gorricho, N.~Bouten, F.~D. Turck, and R.~Boutaba,
  ``Network function virtualization: State-of-the-art and research
  challenges,'' \emph{IEEE Communications Surveys Tutorials}, vol.~18, no.~1,
  pp. 236--262, Firstquarter 2016.

\bibitem{Martins:2014}
J.~Martins, M.~Ahmed, C.~Raiciu, V.~Olteanu, M.~Honda, R.~Bifulco, and
  F.~Huici, ``Clickos and the art of network function virtualization,'' in
  \emph{Proceedings of the 11th USENIX Conference on Networked Systems Design
  and Implementation}, ser. NSDI'14, Berkeley, CA, USA, 2014, pp. 459--473.

\bibitem{Mouradian}
C.~Mouradian, T.~Saha, J.~Sahoo, R.~Glitho, M.~Morrow, and P.~Polakos, ``Nfv
  based gateways for virtualized wireless sensor networks: A case study,'' in
  \emph{IEEE International Conference on Communication Workshop (ICCW)}, June
  2015, pp. 1883--1888.

\bibitem{Munoz}
R.~Muñoz, R.~Vilalta, R.~Casellas, R.~Martínez, T.~Szyrkowiec, A.~Autenrieth,
  V.~López, and D.~López, ``Sdn/nfv orchestration for dynamic deployment of
  virtual sdn controllers as vnf for multi-tenant optical networks,'' in
  \emph{Optical Fiber Communications Conference and Exhibition (OFC)}, March
  2015, pp. 1--3.

\bibitem{Carella}
G.~Carella, M.~Corici, P.~Crosta, P.~Comi, T.~M. Bohnert, A.~A. Corici,
  D.~Vingarzan, and T.~Magedanz, ``Cloudified ip multimedia subsystem (ims) for
  network function virtualization (nfv)-based architectures,'' in \emph{IEEE
  Symposium on Computers and Communications (ISCC)}, vol. Workshops, June 2014,
  pp. 1--6.

\bibitem{Bari}
M.~F. Bari, S.~R. Chowdhury, R.~Ahmed, and R.~Boutaba, ``On orchestrating
  virtual network functions,'' in \emph{11th International Conference on
  Network and Service Management (CNSM)}, November 2015, pp. 50--56.

\bibitem{Jahromi}
N.~Jahromi, S.~Yangui, A.~Larabi, D.~Smith, M.~A. Salahuddin, R.~Glitho,
  R.~Brunner, and H.~Elbiaze, ``Nfv and sdn-based cost-efficient and agile
  value-added video services provisioning in content delivery networks,'' in
  \emph{IEEE Consumer Communications \& Networking Conference (CCNC)}, January
  2017.

\bibitem{ETSI}
``Etsi gs nfv 001, network function virtualization (nfv) use cases, v1.1.1.''
  ETSI, ~, 2013.

\bibitem{Giotis}
K.~Giotis, Y.~Kryftis, and V.~Maglaris, ``Policy-based orchestration of nfv
  services in software-defined networks,'' in \emph{IEEE Conference on Network
  Softwarization (NetSoft)}, April 2015, pp. 1--5.

\bibitem{Herbaut}
N.~Herbaut, D.~Negru, G.~Xilouris, and Y.~Chen, ``Migrating to a nfv-based home
  gateway: Introducing a surrogate vnf approach,'' in \emph{6th International
  Conference on the Network of the Future (NOF)}, September 2015, pp. 1--7.

\bibitem{Luizelli}
M.~C. Luizelli, L.~R. Bays, L.~S. Buriol, M.~P. Barcellos, and L.~P. Gaspary,
  ``Piecing together the nfv provisioning puzzle: Efficient placement and
  chaining of virtual network functions,'' in \emph{IFIP/IEEE International
  Symposium on Integrated Network Management (IM)}, May 2015, pp. 98--106.

\bibitem{Fang}
W.~Fang, M.~Zeng, X.~Liu, W.~Lu, and Z.~Zhu, ``Joint spectrum and it resource
  allocation for efficient vnf service chaining in inter-datacenter elastic
  optical networks,'' \emph{IEEE Communications Letters}, vol.~20, no.~8, pp.
  1539--1542, August 2016.

\bibitem{Moens}
H.~Moens and F.~D. Turck, ``Vnf-p: A model for efficient placement of
  virtualized network functions,'' in \emph{10th International Conference on
  Network and Service Management (CNSM) and Workshop}, November 2014, pp.
  418--423.

\bibitem{Qu}
L.~Qu, C.~Assi, K.~Shaban, and M.~Khabbaz, ``Reliability-aware service
  provisioning in nfv-enabled enterprise datacenter networks,'' in
  \emph{IFIP/IEEE 12th International Conference on Network and Service
  Management}, October 2016.

\bibitem{Xia}
M.~Xia, M.~Shirazipour, Y.~Zhang, H.~Green, and A.~Takacs, ``Network function
  placement for nfv chaining in packet/optical datacenters,'' \emph{Journal of
  Lightwave Technology}, vol.~33, no.~8, pp. 1565--1570, April 2015.

\bibitem{Ghaznavi}
M.~Ghaznavi, A.~Khan, N.~Shahriar, K.~Alsubhi, R.~Ahmed, and R.~Boutaba,
  ``Elastic virtual network function placement,'' in \emph{IEEE 4th
  International Conference on Cloud Networking (CloudNet)}, October 2015, pp.
  255--260.

\bibitem{Ghaznavi2}
{M. Ghaznavi\vspace{0mm} and A. Khan and N. Shahriar and K. Alsubhi and R.
  Ahmed and R. Boutaba}, ``Service function chaining simplified,''
  \url{http://arxiv.org/abs/1601.00751}, coRR, vol. abs/1601.00751, 2016.

\bibitem{Mechtri}
M.~Mechtri, C.~Ghribi, and D.~Zeghlache, ``A scalable algorithm for the
  placement of service function chains,'' \emph{IEEE Transactions on Network
  and Service Management}, vol.~13, no.~3, pp. 533--546, September 2016.

\bibitem{Riggio}
R.~Riggio, T.~Rasheed, and R.~Narayanan, ``Virtual network functions
  orchestration in enterprise wlans,'' in \emph{IFIP/IEEE International
  Symposium on Integrated Network Management (IM)}, May 2015, pp. 1220--1225.

\bibitem{Riggio2}
R.~Riggio, A.~Bradai, D.~Harutyunyan, T.~Rasheed, and T.~Ahmed, ``Scheduling
  wireless virtual networks functions,'' \emph{IEEE Transactions on Network and
  Service Management}, vol.~13, no.~2, pp. 240--252, June 2016.

\bibitem{Sun}
Q.~Sun, P.~Lu, W.~Lu, and Z.~Zhu, ``Forecast-assisted nfv service chain
  deployment based on affiliation-aware vnf placement,'' in \emph{IEEE
  Globecom}, December 2016.

\bibitem{Lin}
T.~Lin, Z.~Zhou, M.~Tornatore, and B.~Mukherjee, ``Demand-aware network
  function placement,'' \emph{Journal of Lightwave Technology}, vol.~34,
  no.~11, pp. 2590--2600, June 2016.

\bibitem{Zeng}
M.~Zeng, W.~Fang, and Z.~Zhu, ``Orchestrating tree-type vnf forwarding graphs
  in inter-dc elastic optical networks,'' \emph{Journal of Lightwave
  Technology}, vol.~34, no.~14, pp. 3330--3341, July 2016.

\bibitem{Bouet}
M.~Bouet, J.~Leguay, and V.~Conan, ``Cost-based placement of vdpi functions in
  nfv infrastructures,'' in \emph{Proceedings of the 1st IEEE Conference on
  Network Softwarization (NetSoft)}, April 2015, pp. 1--9.

\bibitem{Herbaut2017}
N.~Herbaut, D.~Negru, D.~Dietrich, and P.~Papadimitriou, ``Service chain
  modeling and embedding for nfv-based content delivery,'' in \emph{2017 IEEE
  International Conference on Communications (ICC)}, May 2017, pp. 1--7.

\bibitem{Ibn-Kheder2017}
H.~Ibn-Khedher, E.~Abd-Elrahman, A.~E. Kamal, and H.~Afifi, ``Opac: An optimal
  placement algorithm for virtual cdn,'' \emph{Computer Networks}, vol. 120,
  no. Supplement C, pp. 12 -- 27, 2017.

\bibitem{Frangoudis2017}
P.~A. Frangoudis, L.~Yala, and A.~Ksentini, ``Cdn-as-a-service provision over a
  telecom operator's cloud,'' \emph{IEEE Transactions on Network and Service
  Management}, vol.~14, no.~3, pp. 702--716, Sept 2017.

\bibitem{Monaci}
M.~Monaci and P.~Toth, ``A set-covering-based heuristic approach for
  bin-packing problems,'' \emph{INFORMS Journal on Computing}, vol.~18, no.~1,
  pp. 71--85, 2006.

\bibitem{Teixeira200892}
J.~C. Teixeira and A.~P. Antunes, ``A hierarchical location model for public
  facility planning,'' \emph{European Journal of Operational Research}, vol.
  185, no.~1, pp. 92 -- 104, 2008.

\bibitem{Farahani}
R.~Z. Farahani, M.~Hekmatfar, B.~Fahimnia, and N.~Kazemzadeh, ``Survey:
  Hierarchical facility location problem: Models, classifications, techniques,
  and applications,'' \emph{Comput. Ind. Eng.}, vol.~68, pp. 104--117, February
  2014.

\bibitem{BRIN1998107}
S.~Brin and L.~Page, ``The anatomy of a large-scale hypertextual web search
  engine,'' \emph{Computer Networks and ISDN Systems}, vol.~30, no.~1, pp. 107
  -- 117, 1998.

\bibitem{Cheng2011}
X.~Cheng, S.~Su, Z.~Zhang, H.~Wang, F.~Yang, Y.~Luo, and J.~Wang, ``Virtual
  network embedding through topology-aware node ranking,'' \emph{ACM SIGCOMM
  Computer Communication Review}, vol.~41, no.~2, pp. 38--47, 2011.

\bibitem{Son2012}
S.-W. Son, C.~Christensen, P.~Grassberger, and M.~Paczuski, ``Pagerank and
  rank-reversal dependence on the damping factor,'' \emph{Physical Review E},
  vol.~86, no.~6, p. 066104, 2012.

\bibitem{Page1998}
P.~Lawrence, B.~Sergey, R.~Motwani, and T.~Winograd, ``The pagerank citation
  ranking: Bringing order to the web,'' Stanford University, Technical Report,
  1998.

\bibitem{Cacheda2007}
R.~A. Cacheda, D.~C. Garc{\'\i}a, A.~Cuevas, F.~J.~G. Castano, J.~H.
  S{\'a}nchez, G.~Koltsidas, V.~Mancuso, J.~I.~M. Novella, S.~Oh, and
  A.~Pant{\`o}, ``Qos requirements for multimedia services,'' in \emph{Resource
  Management in Satellite Networks}.\hskip 1em plus 0.5em minus 0.4em\relax
  Springer, 2007, pp. 67--94.

\bibitem{Cesar2015}
C.~Ghali, ``Topology generator,''
  \url{https://github.com/cesarghali/topology-generator}, 2015.

\bibitem{Bianchini2005}
M.~Bianchini, M.~Gori, and F.~Scarselli, ``Inside pagerank,'' \emph{ACM
  Transactions on Internet Technology (TOIT)}, vol.~5, no.~1, pp. 92--128,
  2005.

\bibitem{Ahuja1993}
R.~K. Ahuja, T.~L. Magnanti, and J.~B. Orlin, \emph{Network flows: theory,
  algorithms, and applications}.\hskip 1em plus 0.5em minus 0.4em\relax
  Prentice hall, 1993.

\end{thebibliography}
\vskip -1.78\baselineskip
\begin{biography}[
{\includegraphics[width=1in,height=1.25in,clip,keepaspectratio]{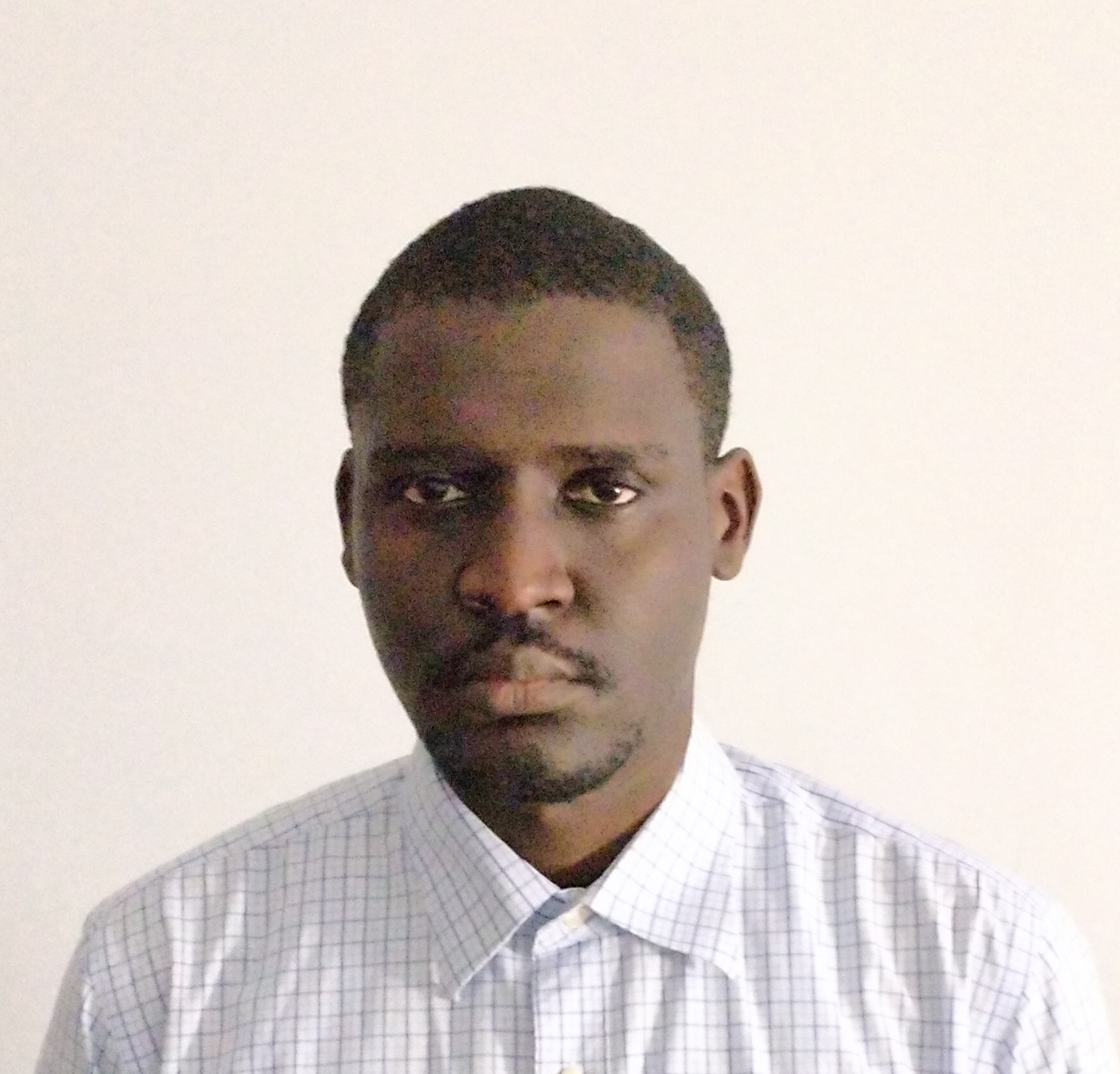}}
]
{MOUHAMAD DIEYE} received the Engineering degree in teleinformatics from the Ecole Superieure Multinationale des Telecommunications in collaboration with the Ecole Superieure Polytechnique, Senegal in 2012, and the M.Sc. degree in computer science from the Université du Québec à Montréal where he is currently pursuing a Ph.D candidate. His research interests cloud computing and storage, replication strategies, storage techniques, network function virtualization and fog
computing. He also currently serves as a reviewer for several international conferences and journals.
\end{biography}
\vskip 0pt plus -1fil
\begin{IEEEbiography}[{\includegraphics[width=1in,height=1.25in,clip,keepaspectratio]{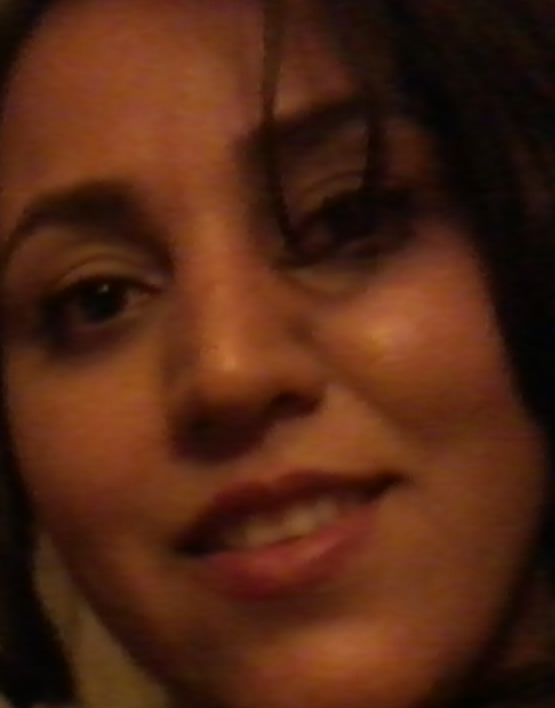}}]
{SHOHREH AHVAR} received B.S degree in Information Technology and M.S degree in Electrical Engineering-Telecommunication (Networks) both from Isfahan University of Technology, Isfahan, Iran. She is currently a Ph.D candidate at the Institut Mines-Telecom, Telecom SudParis in co-accreditation with the Pierre and Marie Curie University (Paris 6) on the topic of cloud based content delivery networks. She has participated in the FUSE-IT ITEA project for 3 years managed by Airbus and she is also a collaborating researcher at Concordia University, Montreal, Canada working on Ericson funded project. Her research interests are Network Function Virtualization, Content Delivery Networks, Cloud Computing, Wireless Sensor Networks and Building Management Systems.
\end{IEEEbiography}
\vskip 0pt plus -1fil
\begin{IEEEbiography}[{\includegraphics[width=1in,height=1.25in,clip,keepaspectratio]{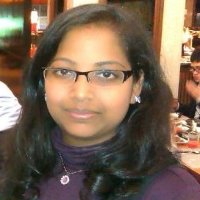}}]
{JAGRUTI SAHOO} received the Ph.D. degree in computer science and information engineering from the National Central University, Taiwan, 2013. She has been a Post-Doctoral Fellow with the University of Sherbrooke,Canada, and Concordia University, Canada. She is currently an Assistant Professor with the Department of Mathematics and Computer Science, South Carolina State University, USA. Her research interests include wireless sensor networks, vehicular networks, content delivery networks, cloud computing, and network functions virtualizations. She has served as a member of the Technical Program Committee of many conferences and a reviewer of many journals and conferences.
\end{IEEEbiography}
\vskip 0pt plus -1fil
\begin{IEEEbiography}[{\includegraphics[width=1in,height=1.25in,clip,keepaspectratio]{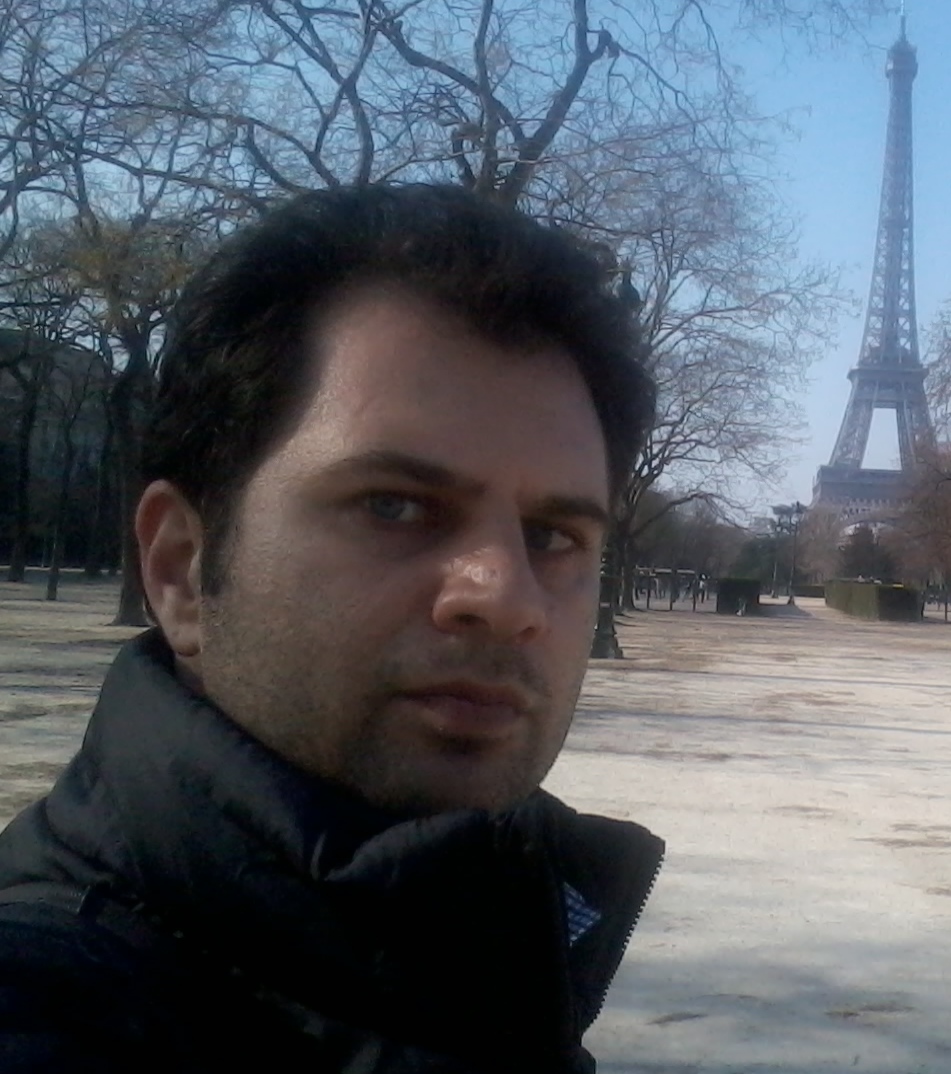}}]
{EHSAN AHVAR} received a Phd degree in computer science and telecommunications from Telecom SudParis and Paris.VI  University  (co-accreditation), France. He was a faculty member of Information Technology department at Payame Noor University (P.N.U), Iran from 2007 to 2012. He is currently a post doctoral researcher at Inria Rennes-Bretagne Atlantique, France. His research  interests include cloud computing, network functions virtualization (NFV), content delivery networks (CDN), wireless sensor networks and Internet of Things.
\end{IEEEbiography}
\vskip 0pt plus -1fil
\begin{IEEEbiography}[{\includegraphics[width=1in,height=1.25in,clip,keepaspectratio]{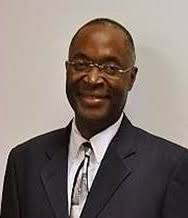}}]
{Roch Glitho} (M’88–SM’97) received the Ph.D.
(Tekn.Dr.) degree in teleinformatics from the Royal
Institute of Technology, Stockholm, Sweden, the
M.Sc. degree in business economics from the
University of Grenoble, France, and the M.Sc.
degree in pure mathematics and the M.Sc. degree
in computer science from the University of Geneva,
Switzerland. He is an Associate Professor and
a Canada Research Chair with Concordia University.
He is also an Adjunct Professor at several other universities,
including Telecom Sud Paris, France, and
the University of Western Cape, South Africa. He has worked in industry
and has held several senior technical positions, including a Senior Specialist,
a Principal Engineer, an Expert with Ericsson in Sweden and Canada.
His industrial experience includes research, international standards setting,
product management, project management, systems engineering, and software/firmware
design. He has also served as an IEEE Distinguished Lecturer,
and the Editor-in-Chief of the IEEE Communications Magazine and the IEEE
COMMUNICATIONS SURVEYS \& TUTORIALS JOURNAL.
\end{IEEEbiography}
\vskip 0pt plus -1fil
\begin{IEEEbiography}[{\includegraphics[width=1in,height=1.25in,clip,keepaspectratio]{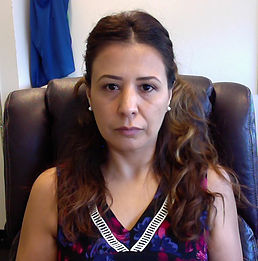}}]
{HALIMA ELBIAZE (M'06)} received the B.Sc. degree in applied mathematics from the University
of MV, Morocco, in 1996, the M.Sc. degree in telecommunication systems from the Université de
Versailles in 1998, and the Ph.D. degree in computer science from the Institut National des Télé-communications, Paris, France, in 2002. She is currently an Associate Professor with the Department of Computer Science, Université du Québec à Montréal, QC, Canada, where has been serving since 2003. She has authored and co-authored many journal and conference papers. Her research interests include network performance evaluation, traffic engineering, cloud computing, and quality of service management in optical and wireless networks.
\end{IEEEbiography}
\vskip 0pt plus -1fil
\begin{IEEEbiography}[{\includegraphics[width=1in,height=1.25in,clip,keepaspectratio]{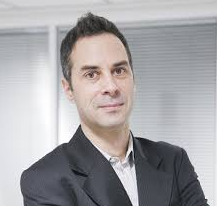}}]
{NOEL CRESPI}  received the master’s degrees from
the University of Orsay (Paris 11) and the University
of Kent, U.K., the Diplome d’Ingenieur
degree from Telecom ParisTech, and the Ph.D. and
Habilitation degrees from Paris VI University
(Paris-Sorbonne). Since 1993, he has been with
CLIP, Bouygues Telecom, and then with Orange
Labs in 1995. He took leading roles in the creation
of new services with the successful conception and
launch of Orange prepaid service, and in standardization
(from rapporteurship of IN standard to coordination of all mobile
standards activities for Orange). In 1999, he joined Nortel Networks as a
Telephony Program Manager, architecting core network products for EMEA
region. In 2002, he joined the Institut Mines-Telecom and is currently a
Professor and the Program Director, leading the Service Architecture Laboratory.
He coordinates the standardization activities for the Institut MinesTelecom,
ITU-T, ETSI, and 3GPP. He is an Adjunct Professor with KAIST,
an Affiliate Professor with Concordia University, and a Guest Researcher
with the University of Goettingen. He is also the Scientific Director of the
French-Korean Laboratory ILLUMINE. His current research interests are in
sofwarization, data analysis, and Internet of Things/Services.
\end{IEEEbiography}
\end{document}